\documentclass[%
 aip,
amsmath,amssymb,
reprint,%
]{revtex4-1}

\usepackage{graphicx}
\usepackage{dcolumn}
\usepackage{bm}
\usepackage{color}
\usepackage{ulem}

\usepackage[utf8]{inputenc}
\usepackage[T1]{fontenc}
\usepackage{mathptmx}

\usepackage{hyperref}
\hypersetup{hypertex=true,
	colorlinks=true,
	linkcolor=blue,
	anchorcolor=blue,
	citecolor=blue}

\begin{document}

\preprint{AIP/123-QED}

\title{Polarized exciton emission enhancement of monolayer MoS$_2$ coupled with plasmonic Salisbury-type absorber}

\author{Wei Li}
\affiliation{Beijing National Laboratory for Condensed Matter Physics, Institute of Physics, Chinese Academy of Sciences, Beijing 100190, P.R. China}
\affiliation
{School of Physical Sciences, CAS Key Laboratory of Vacuum Physics, University of Chinese Academy of Sciences, Beijing 100190, P.R. China}

\author{Ming Xin}
\affiliation{Beijing National Laboratory for Condensed Matter Physics, Institute of Physics, Chinese Academy of Sciences, Beijing 100190, P.R. China}
\affiliation
{School of Physical Sciences, CAS Key Laboratory of Vacuum Physics, University of Chinese Academy of Sciences, Beijing 100190, P.R. China}

\author{Wenze Lan}
\affiliation{Beijing National Laboratory for Condensed Matter Physics, Institute of Physics, Chinese Academy of Sciences, Beijing 100190, P.R. China}
\affiliation
{School of Physical Sciences, CAS Key Laboratory of Vacuum Physics, University of Chinese Academy of Sciences, Beijing 100190, P.R. China}

\author{Qinghu Bai}
\affiliation{Beijing National Laboratory for Condensed Matter Physics, Institute of Physics, Chinese Academy of Sciences, Beijing 100190, P.R. China}
\affiliation
{School of Physical Sciences, CAS Key Laboratory of Vacuum Physics, University of Chinese Academy of Sciences, Beijing 100190, P.R. China}

\author{Shuo Du}
\affiliation{Beijing National Laboratory for Condensed Matter Physics, Institute of Physics, Chinese Academy of Sciences, Beijing 100190, P.R. China}
\affiliation
{School of Physical Sciences, CAS Key Laboratory of Vacuum Physics, University of Chinese Academy of Sciences, Beijing 100190, P.R. China}

\author{Gang Wang}
\affiliation{School of Physics, Beijing Institute of Technology, Beijing100081, P.R. China.}

\author{Baoli Liu}
\email{blliu@iphy.ac.cn}
\affiliation{Beijing National Laboratory for Condensed Matter Physics, Institute of Physics, Chinese Academy of Sciences, Beijing 100190, P.R. China}
\affiliation
{CAS Center for Excellence in Topological Quantum Computation, CAS Key Laboratory of Vacuum Physics, University of Chinese Academy of Sciences, Beijing 100190, P.R. China}
\affiliation
{Songshan Lake Materials Laboratory, Dongguan, Guangdong 523808, P.R. China}

\author{Changzhi Gu}
\email{czgu@iphy.ac.cn}
\affiliation{Beijing National Laboratory for Condensed Matter Physics, Institute of Physics, Chinese Academy of Sciences, Beijing 100190, P.R. China}
\affiliation
{School of Physical Sciences, CAS Key Laboratory of Vacuum Physics, University of Chinese Academy of Sciences, Beijing 100190, P.R. China}

\date{\today}

\begin{abstract}
The plasmon-mediated manipulation of light-matter interaction in two-dimensional atomically transition-metal dichalcogenides (TMDs) critically depends on the design of plasmonic nanostructures to achieve the maximum optical field in TMDs. Here, a metal-isolator-metal Salisbury-type perfect absorber was fabricated to serve as a generator of the localized surface plasmons. The significant photoluminescence (PL) enhancement up to 60-fold was observed experimentally in the monolayer (ML) MoS$_{2}$ on the top of this gold plasmonic hybrid nanostructures. Furthermore, the PL linear polarization can approach $\sim$ 60 $\%$ around the peak of exciton emission and is independent on the polarization of the excitation laser. This Salisbury-type plasmon-exciton hybrid system paves a new way to develop optoelectronic devices based on TMDs.

	~\\
	~\\
\end{abstract}

\maketitle

The atomically thin transition-metal dichalcogenides (TMDs) have drawn extensive attention owing to their excellent performance in the field of optoelectronics\cite{Humberto2013,Zeng2012,Yu2014,Lu2015}, mechanics,\cite{Hiram2013,Castellanos2013} and so forth. In contrast to their bulk counterparts, the monolayer (ML) TMDs exhibit much stronger exciton emission in the visible and near-infrared regions stemming from the electronic band structures of the direct bandgap.\cite{Mak2010,Splendiani2010,Jin2013} The excitons of TMD monolayers are also extremely stable at room temperature owing to the large exciton binding energy up to several hundred meV.\cite{Chernikov2014,Hanbicki2015,Zhu2015,He2014} As a promising material, many great device applications in the field of optoelectronics have been realized experimentally such as photodetectors,\cite{Lei2019,Juan2019} solar cells,\cite{Feifei2018} chemical sensors\cite{Gerkman2020} based on the monolayer TMDs. Particularly, the broken inversion symmetry and strong spin-orbit interactions yield the well-known spin-valley properties revealed by the chiral optical selection rule of valley exciton in those MLs.\cite{Cao2012,Zeng2012,Mak2012,Xiao2012} The coherent superposition of K and -K valley exciton states, as a quantum information carrier, can be generated by linearly polarized light and characterized by the linearly polarized photoluminescence (PL) emission to a certain extent in low temperature.\cite{Jones2013,Wang2016,Bouet2014,Glazov2015} However, their application in photonic and valleytronic devices is limited by their low quantum efficiency and fast quantum decoherence,\cite{Li2016,Wang2016} respectively. Fortunately, the plasmonic nanostructures can be integrated with TMDs to enhance the quantum efficiency\cite{Torma2015} and suppress the quantum decoherence\cite{Bogdanov2019} by the enhancement of exciton-plasmon coupling owing to their extraordinary capability of localizing electromagnetic field within the subwavelength scale.\cite{Kuttge2010,Akselrod2014,Monticone2017}
 Recently, many hybrid TMDs-plasmonic systems, including metallic nanoparticle arrays, nanoslits, and nanoarrays, were widely explored to investigate the exciton-plasmon coupling characterized by the enhancement and the polarization control of PL emission. \cite{Li2018,Peinan2019,Han2020,Kim2020,Chervy2018,Gong2018,Sun2019,Deng2020,Qi2020} In principle, the optically bright exciton of TMDs is composed of the conduction and valence electronic states with the same spin projections in each valley. The out-of-plane emission of bright excitons is characterized by strong in-plane optical dipoles
 ,\cite{Wang2016,Wang2018,Glazov2014} which can couple to the optical field with an in-plane electric vector. Normally, the metal nanostructures upon interacting with light produce collective charge oscillations, known as localized surface plasmons (LSPs), which induce a strong resonant optical field in their vicinity. In this regard, it is highly desired to look for an optimal plasmonic hybrid nanostructure, in which the stronger in-plane optical field can be achieved.
 
\begin{figure}
	\includegraphics[width=0.45\textwidth]{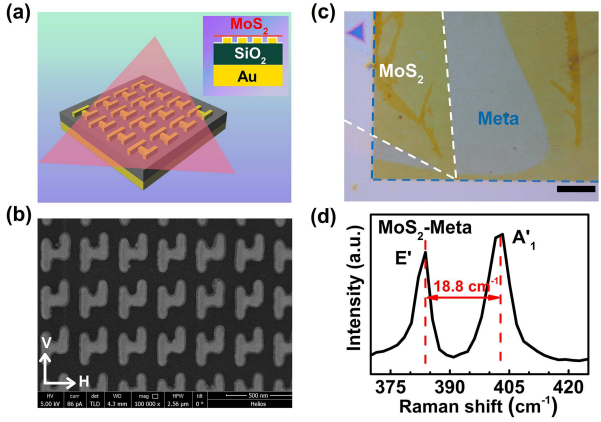}
	\caption{(a) 3D schematics of the hybrid nanostructure. From bottom to top: gold mirror, SiO$_2$ insulating layer, gold metasurface, Al$_2$O$_3$ insulating layer, and CVD-grown ML MoS$_2$. (b) SEM image of the gold metasurface fabricated by EBL and EBE. (c) The optical microscopic image of the hybrid nanostructure. The scale bar is 20 $\mu m$. (d) Raman spectra in the MoS$_2$-metasurface.}
	\label{Figure 1}
\end{figure}

In this work, we designed a ML MoS$_{2}$/gold plasmonic hybrid nanostructure which consists of a metal-isolator-metal (MIM) based Salisbury-type perfect absorber. Thanks to the unique property of the maximum optical field of Salisbury-type absorber,\cite{Fante1988,Ra'di2015} in which a layer of the metal metasurface is located at quarter-wavelength distance from the metal ground. We found that the integrated PL intensity of ML MoS$_{2}$ exhibits a 60-fold enhancement and the linear polarization of PL is up to $\sim$ 60 $\%$ around the peak of exciton emission. Moreover, the linear polarization of light emitting is independent on the direction of the electric field of excitation light. Our hybrid Salisbury-type plasmon-enhanced design provides full access to the flexible manipulation of PL emission and its polarization for integrated and ultrathin optoelectronic and valleytronic devices.

To accomplish the efficient PL enhancement and polarization engineering, the hybrid system consisting of 2H-monolayer MoS$_2$ and a MIM-based Salisbury-type perfect absorber was fabricated, as illustrated in Figure \ref{Figure 1}(a). As a MIM-based perfect absorber, the 100 nm gold layer at the bottom serves as a reflecting mirror, and the 140 nm amorphous quartz was deposited on the gold mirror. The periodic gold metasurface was then fabricated in a series of micro-nano processing steps (see Methods). Afterward, the 5 nm Al$_2$O$_3$ was deposited onto the metasurface. Finally, the monolayer MoS$_2$ prepared by the chemical vapor deposition (CVD) method was transferred on the top of the metasurface to wipe off the polarization selectivity of the excitation laser and the light emitting of excitons passing through the metasurface. In addition, a thin layer of 5 nm Al$_2$O$_3$ beneath the monolayer MoS$_2$ was employed to restrain the hot-electron transfer process between metasurface and MoS$_2$ which will result in the quenching of exciton emission. Figure \ref{Figure 1}(b) shows the scanning electron microscopy (SEM) image of the gold metasurface with a scale bar of 500 nm. The h-shaped unit, which consists of three gold nanorods, is chosen for the metasurface to get the anisotropy of plasmonic resonant absorption. The period of the unit is $\sim$300 nm along the H-axis and $\sim$360 nm along the V-axis, respectively. The width and the height of the nanorods are 50 nm. The length of each nanorod is 250 nm, 110 nm, and 190 nm, respectively. 

Figure \ref{Figure 1}(c) presents the optical microscope image of the hybrid sample. The light green part on the metasurface (square pattern) is the transferred monolayer MoS$_2$ characterized by the standard Raman spectrum. It is clear that the in-plane vibration mode ($\mathrm{E'}$) and the out-of-plane vibration mode ($\mathrm{A'_1}$) can be observed, as shown in Figure \ref{Figure 1}(d). And then, the wavenumber difference of the $\mathrm{E'}$ mode and the $\mathrm{A'_1}$ mode is about $\sim$18.8 cm$^{-1}$. This notable feature of monolayer MoS$_2$ is consistent with the previous experimental observation.\cite{Lee2010}

\begin{figure}
	\includegraphics[width=0.45\textwidth]{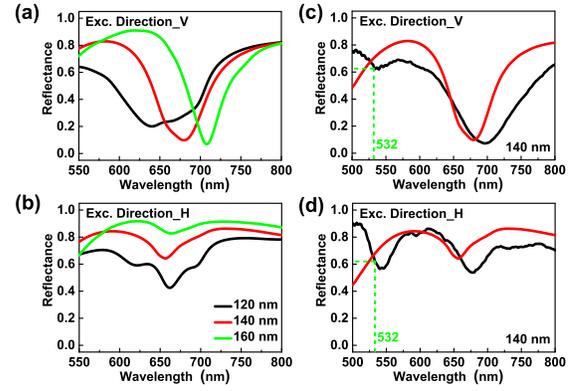}
	\caption{Simulated reflectance spectra of metasurface with different insulating layer thickness for the polarization of excitation light along (a) V-axis and (b) H-axis. (c)-(d) Experimental (black solid curves) and  simulated (red solid curves) reflectance spectra of metasurface with a 140nm isolator layer for orthogonal polarizations.}
	\label{Figure 2}
\end{figure}

In order to evaluate the optical responses of the designed Salisbury-type perfect absorber with a symmetry-breaking metasurface, we simulated the reflectance spectra of this plasmonic generator varying the thickness of Silica isolator layer from $\sim$ 120 nm to $\sim$ 160 nm with the orthogonal polarizations of broadband incident white light as shown in Figure \ref{Figure 2}(a) and (b). Three distinct features were obtained: 1) the deeper dips in those spectra were gotten with the polarization of incident light along the V-axis compared to that along the H-axis for three kinds of Silica isolator layer thicknesses, which correspond to the resonant absorption of plasmons; 2) for a given thickness of isolator layer, the resonant wavelength of plasmons is obviously different for orthogonal polarizations. The red shift of resonant wavelength was observed with the polarization along V-axis for the layer thicknesses of $\sim$ 140 nm and $\sim$ 160 nm; 3) for a 140 nm layer thickness, the resonant wavelength of $\lambda_{0} \sim$ 680 nm perfectly matches the excitonic optical transition of our CVD-grown ML MoS$_{2}$ as we expect. Furthermore, this thickness of $\sim$ 140 nm is almost equal to $\lambda_{0}$/4, which fulfilled the characterized requirement of Salisbury-type perfect absorber, with the consideration of the effective refractive index of Silica plus metasurface. Figure \ref{Figure 2}(c) and (d) present the experimental measurements of the reflectance spectra with the orthogonal polarizations on the fabricated plasmonic Salisbury-type perfect absorber as indicated by black solid curves. It is found that those two spectra were almost consistent with the simulated results of $\sim$ 140 nm isolator layer thickness although there exists a small deviation due to the imperfect fabrication.

\begin{figure}
	\includegraphics[width=0.45\textwidth]{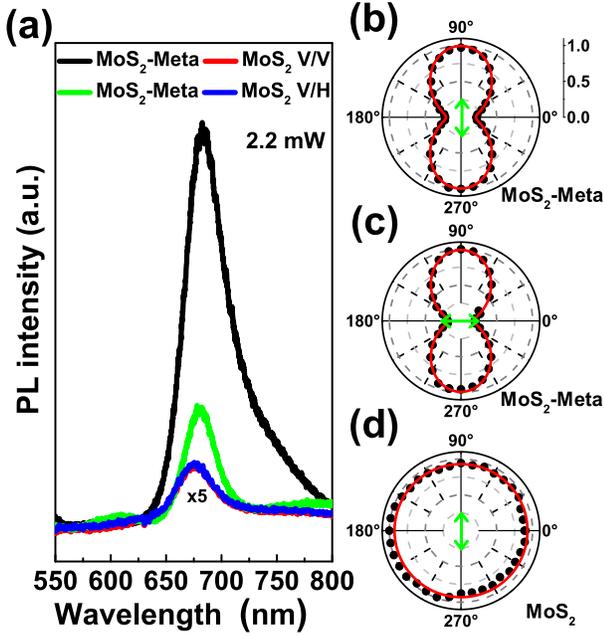}
	\caption{(a) Linear polarization-resolved photoluminescence spectra for vertically (V, black and red curves) and horizontally (H, green and blue curves) polarized detection. The incident laser is vertically polarized. (b) Normalized A-exciton peak intensity as a function of detection angle for given incident laser polarization (marked by green arrow). 0 corresponds to the centre and 1.0 to the outermost dashed circle.}
	\label{Figure 3}
\end{figure}

To reveal the effect of the gold plasmonic metasurface with the symmetry-breaking h-shape units on the exciton emission of ML MoS$_2$, we performed the measurements of the linear polarization resolved PL spectra of ML MoS$_2$-metasurface and the bare ML MoS$_2$ at room temperature with the polarizations of excitation laser alone V-axis and a fixed power of $\sim$2.2 mW as shown in Figure \ref{Figure 3}(a) (see Optical measurement of Methods). The featured A-exciton light emission of ML MoS$_2$ was observed around $\sim$ 680 nm for both samples. It is obvious that the PL intensity of the MoS$_2$-metasurface is stronger than that of the bare ML MoS$_2$ for both of V- and H-polarization components. More interesting, the V-polarization component of the PL intensity is stronger than H-polarization component for the sample of ML MoS$_2$-metasurface while both components of the PL intensities remain the same for the bare ML MoS$_2$. In order to obtain the information of polarization direction of maximum PL intensity for ML MoS$_2$-metasurface, we measured the detection-angle dependent peak intensity of PL spectra with the incident laser polarization along V-axis as shown in Figure \ref{Figure 3}(b). It is clear that the polarization direction of maximum PL intensity is along V-axis. This result is consistent with the simulated reflectance spectra as presented in Figure \ref{Figure 2}(a) and (b), in which the maximum resonant absorption of plasmons occurred with polarization of incident light along V-axis. The polarization anisotropy can be characterized by the degree of linear polarization $P_{lin.}=\frac{I_{V}-I_{H}}{I_{V}+I_{H}}$, leading to the degree of the linear polarization around $\sim 60\%$ at the main peak of exciton emission. Furthermore, the polarization of maximum PL intensity is still along V-axis and $P_{lin.}$ remains unchanged when the polarization of the excitation laser is parallel to H-axis . This experimental observation can be manifested more clearly by the polar plot of the polarization-dependent PL intensities in Figure \ref{Figure 3}(c), in which the green arrow denotes the polarization of the excitation laser. On the contrary, there is no any linear polarization of emitting light for the bare ML MoS$_2$ due to the rapid decoherence decay of valley excitons as shown by the polarization-dependent PL intensities in Figure \ref{Figure 3}(d). It is worth noticing that the PL intensity with the polarization along V-axis is the same for both polarization directions of excitation laser with fixed pump power. This feature implied that the effect of the gold plasmonic metasurface on the transition rate of excitons is identical at the excitation wavelength of $\sim$ 532 nm with different laser polarization, which is supported by the same plasmon absorptions of metasurface at this wavelength of $\sim$ 532 nm as shown in Figure \ref{Figure 2}(c) and (d). The linear polarization of emitting light in ML TMDs corresponds to the coherent superposition of valley excitons.\cite{Wang2016} The strong anisotropic characterization of linearly polarized exciton emission illustrated that the Salibury-based plasmonic nanostructure provides an effective route to manipulate the linear polarization  of TMDs in the field of valleytronics.

\begin{figure}
	\includegraphics[width=0.45\textwidth]{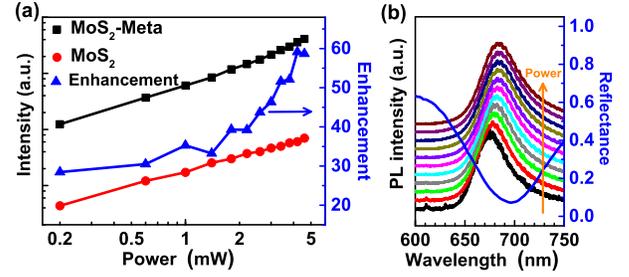}
	\caption{(a) PL enhancement factor and integrated PL intensities as a function of excitation powers. Incident laser and PL are both vertically polarized. (b) PL spectra of MoS$_2$-metasurface sample excited with laser power ranging from 0.2 to 4.6 mW. The blue curve shows the reflectance spectra of the plasmonic nanostructure.}
	\label{Figure 4}
\end{figure}

Now, we turn to analyze the enhancement effect of PL intensity induced by metasurface with the polarization along V-axis (See Supporting Information for the polarization along H-axis). It is found that this enhancement factor of integrated PL intensity is over 35 folds with the excitation power of $\sim$2.2 mW and significantly larger than that of former reports using similar MIM-based plasmonic nanostructures.\cite{Li2018,Deng2020,Qi2020} In principle, the enhancement of PL intensity in MoS$_2$-metasurface implied the speed-up radiative recombination rate of ML MoS$_2$ A-excitons via the Purcell effect\cite{Purcell1946} induced by the plasmon resonance of metasurface in MoS$_2$-metasurface comparing to that in bare ML MoS$_2$. This postulation can be confirmed by the power-dependent measurements of the PL spectra for both samples. In ML TMDs, there exists the effect of the exciton-exciton annihilation \cite{Sun2014,Froehlicher2016,Wang2019} which strongly depends on the density of photogenerated excitons and will result in the luminescence quenching of exciton emission with increasing the excitation power. The mutual competition of the radiative recombination and the exciton-exciton annihilation will lead to the totally different power-dependent PL intensity of exciton emission. Figure \ref{Figure 4}(a) presents the enhancement factor (blue curve) and the integrated PL intensities (black and red curves) as a function of excitation powers ranging from $\sim$0.2 mW to $\sim$4.6 mW at room temperature. For the enhancement factor, it is around $\sim$30 with the excitation power less than $\sim$1.0 mW and rapidly rises to $\sim$60 with maximum excitation power of $\sim$ 4.6mW. This means the behaviours of power-dependent integrated PL intensities are totally different for the samples of MoS$_2$-metasurface and bare ML MoS$_2$. For the sake of simplicity to clarify this difference, the power-dependent integrated PL intensities of the MoS$_2$-metasurface and ML MoS$_2$ is described by the power law expression: $I_{PL} \propto P^{\beta}$, where $I_{PL}$ is the integrated PL intensity, $P$ is the excitation power of laser. The beta coefficients are $\sim$1.12 and $\sim$0.87 for the MoS$_2$-metasurface and the bare ML MoS$_2$ through the linear fitting in logarithmic coordinates. Those results suggest that the effective radiative recombination of excitons is dominant and the exciton-exciton annihilation can be ignored in the MoS$_2$-metasurface hybrid system (see Supporting Information), which is a sold evidence of the enhancement effect of plasmon resonance on exciton emission due to the significantly accelerated radiative recombination rate. On the contrary, the exciton-exciton annihilation plays a role in the bare ML MoS$_2$. It is worth noticing that, as presented in Figure \ref{Figure 4}(b), there is a small deviation between the PL peak of exciton emission for MoS$_2$-metasurface hybrid system and the maximum absorption of plasmon resonance. Meanwhile, the PL peaks of exciton emission exhibit distinct red shifts with the increasing excitation powers owing to the bandgap renormalization effect.\cite{Park2017} Therefore, the strengthened exciton-plasmon coupling, which will result in the nonlinear power-dependent PL intensity, can be expected in the MoS$_2$-metasurface. As a result, the PL enhancement of $\sim$60 folds can be confirmed at least in our hybrid system.

In conclusion, a metal-isolator-metal Salisbury-type perfect absorber with the symmetry-breaking metasurface was successfully designed and demonstrated to realize the manipulation of light emission and its linear polarization through the integration with monolayer MoS$_2$, benefitting from its unique property of maximum in-plane optical field at plasmon resonance in the vicinity of the metasurface. Due to the feature of evanescent field induced by plasmon resonance, we expect that further PL enhancement and polarization control can be achieved through the replacement of Al$_2$O$_3$ isolator by a layer of more flat and high-K hexagonal Boron Nitride between ML MoS$_2$ and gold metasurface. Our delicate design of hybrid exciton-plasmon system based on Salisbury-type perfect absorber paves a way to develop the novel optoelectronics and valleytronics devices with TMDs.

\textbf{Methods} \textsl{Material Growth.} Molybdenum disulfide monolayers were synthesized by the chemical vapor deposition (CVD) method. Molybdenum trioxide (MoO$_3$) and sulfur (S) powders were served as a precursor, glass chip as substrate, and argon as carrying gas. Firstly, the quartz tube was flushed by a high flow rate of high-purity argon (200 sccm) for 20 min to remove the air completely. Then the argon flow rate was set as 80 sccm, and the furnace was rapidly heated to 800 $^{\circ}$C for MoO$_3$ at 15$^{\circ}$C/min and a separate heater heats the sulfur source to 150 $^{\circ}$C, yielding triangle monolayer domains. The continuous heating process lasts for 10min, followed by natural cooling down to room temperature.

\textsl{Device Fabrication.} The gold reflection layer and the SiO$_2$ insulating layer were deposited by electron beam evaporation (EBE) and plasma enhanced chemical vapor deposition (PECVD), respectively. The periodic gold metasurface structure was fabricated through electron beam lithography (EBL), electron beam evaporation, and lift-off process. 200nm photoresist (PMMA495-A5) was spin coated before electron beam lithography. 5nm titanium (adhesion layer) and 50nm gold were grown by electron beam evaporation after EBL. The sample was immersed in acetone during the lift-off process. 5nm Al$_2$O$_3$ was deposited onto the surface of metasurface via atomic layer deposition system (ALD) at the temperature of 250 $^{\circ}$C. Finally, the CVD grown MoS$_2$ was transferred onto the metasurface via the wetting transfer method.

\textsl{Optical Measurement.} The polarization-resolved micro-Raman/PL spectra were performed in a backscattering geometry using a Jobin-Yvon iHR550 Spectrometer equipped with a liquid-nitrogen-cooled charge-coupled device (CCD). The 300-lines/mm and 600-lines/mm grating were used for PL and Raman measurements, respectively. The optical excitation wavelength is 532 nm from a Helium-Cadmium laser. For the measurements of the linearly resolved PL spectra, the polarization of the incident light was controlled by a Glan-Thompson linear polarizer (Newports). The polarization of incident light was set to be vertical and horizontal, and the polarization of the emission PL was analyzed by rotating the analyzer. We used a long-working-distance 50$\times$ objective (Nikon) with a numerical aperture equal to 0.40 for optical measurement at room temperature. The spectra of reflectance contrast were measured with an ultrafast fiber laser source (SC400-4 Fianium) by normalizing the spectra from the sample on the substrate to that from the bare substrate.

\section*{ASSOCIATED CONTENT}

\textbf{Supporting Information}

Power-dependent PL enhancement factor and integrated PL intensities with polarization of laser along H-axis; formulas derivation of power-dependent PL intensity.

\section*{AUTHOR INFORMATION}

\textbf{Corresponding Author}
~\\*E-mail: blliu@iphy.ac.cn, 
~\\*E-mail: czgu@iphy.c.cn
~\\

\textbf{Author Contributions}

W.L. and M.X. contributed equally to this work. 
~\\

\textbf{Notes}

The authors declare no competing financial interest.

\section*{ACKNOWLEDGMENTS}

This work is supported by the National Natural Science Foundation of China (Grant No. 11974386, 61888102, 61390503, 11874405, 11904019, 12074033, 11574357 ), the National Key Research Program of China (Grant No. 2016YFA0300601), the Strategic Priority Research Program of Chinese Academy of Sciences (Grant No. XDB28000000), the Key Research Program of the Chinese Academy of Sciences (Grant No. XDPB22), the National Basic Research Program of China (Grant No. 2015CB921001), and Beijing Natural Science Foundation (Grant No. Z190006).

\bibliography{reference}

\newpage
\section*{For Table of Contents Only}
	\includegraphics{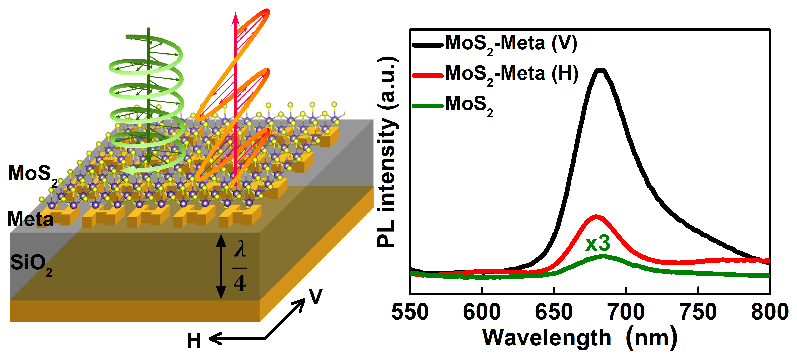}
	\centering 
	\label{TOC}

\end{document}